\documentclass[twocolumn]{jpsj2} 

\title{$^{31}$P-NMR and $\mu$SR Studies of Filled Skutterudite Compound SmFe$_4$P$_{12}$:\\
Evidence for Heavy Fermion Behavior with Ferromagnetic Ground State}

\author{Kenichi \textsc{Hachitani}$^{1,2}$\thanks{E-mail address: hachitani@physics.s.chiba-u.ac.jp},
Hideto \textsc{Fukazawa}$^{1,3}$,
Yoh \textsc{Kohori}$^{1,3}$,
Isao \textsc{Watanabe}$^{2}$,\\
Yuichi \textsc{Yoshimitsu}$^{4}$,
Ken-ichi \textsc{Kumagai}$^{4}$,
Ram \textsc{Giri}$^{5}$,
Chihiro \textsc{Sekine}$^{5}$
and
Ichimin \textsc{Shirotani}$^{5}$}

\inst{$^{1}$Graduate School of Science and Technology, Chiba University, Chiba 263-8522 \\
$^{2}$Advanced Meson Science Laboratory, RIKEN (The Institute of Physical and Chemical Research), Wako, Saitama 351-0198 \\
$^{3}$Department of Physics, Faculty of Science, Chiba University, Chiba 263-8522 \\
$^{4}$Division of Physics, Graduate School of Science, Hokkaido University, Sapporo 060-0810 \\
$^{5}$Department of Electrical and Electronic Engineering, Muroran Institute of Technology, Muroran, Hokkaido 050-8585}

\abst{
The $^{31}$P-NMR (nuclear magnetic resonance) and $\mu$SR (muon spin relaxation) measurements
on the filled skutterudite system SmFe$_4$P$_{12}$ have been carried out.
The temperature $T$ dependence of the $^{31}$P-NMR spectra indicates
the existence of the crystalline electric field effect splitting
of the Sm$^{3+}$ ($J$ = 5/2) multiplet into a ground state and an excited state of about 70~K.
The spin-lattice relaxation rate 1/$T_1$ shows the typical behavior of the Kondo system,
$i.e.$, 1/$T_1$ is nearly $T$ independent above 30 K,
and varies in proportion to $T$ (the Korringa behavior, $1/T_1 \propto T$) between 7.5 K and 30 K.
The $T$ dependence deviated from the Korringa behavior below 7 K,
which is independent of $T$ in the applied magnetic field of 1 kOe,
and suppressed strongly in higher fields.
The behavior is explained as
$1/T_1$ is determined by ferromagnetic fluctuations of the uncovered Sm$^{3+}$ magnetic moments
by conduction electrons.
The $\mu$SR measurements in zero field show
the appearance of a static internal field associated with the ferromagnetic order below 1.6 K.
}

\kword{
SmFe$_4$P$_{12}$, filled skutterudite, ferromagnetism, heavy fermion system, $^{31}$P-NMR (nuclear magnetic resonance) , $\mu$SR (muon spin relaxation)}

\begin{document}
\maketitle

\section{Introduction} 
Filled skutterudite compounds $RT_4X_{12}$ ($R$: rare earth, actinide; $T$: Fe, Ru, Os; $X$: P, As, Sb) 
crystallize in a body-centered cubic (BCC) structure of the space group \textit{Im}~$\bar{3}$ (No.~204).
\cite{ACSB333401}
These compounds have recently attracted much attention as improved thermoelectric materials
and for their wide variety of electrical and magnetic properties,
such as
the superconductivity,
the heavy fermion (HF) system,
the multipolar order,
the magnetic order,
the metal-insulator (M-I) transition
and the non-Fermi-liquid behavior.
\cite{PRB567866,JPCM106973,PRB368660,PRL793218,JPCM135971,JPSJ742141,PRB73052408,JPSJ75}

Among them, SmRu$_4$P$_{12}$ was reported to exhibit an M-I transition at $T_{\text{MI}}$ of 16.5~K.
Recently, it was demonstrated that the M-I transition accompanies the magnetic octupolar order
and that there is the additional antiferromagnetic (AFM) order below 15~K, respectively.
\cite{CTHP2000826,JPSJ71Sp237,APP34983,JPSJ742141,PRB73052408,JPSJ75}
The SmOs$_4$P$_{12}$ was reported to show the simple AFM order below 4.6~K. \cite{PB329458,PB-Hachitani}
The SmFe$_4$P$_{12}$ and SmOs$_4$Sb$_{12}$ were reported to
exhibit the heavy fermion behavior with ferromagnetic (FM) ground states.
\cite{PB329458,JPCM15229,IPAP_CS5_37_2004,PB-Hachitani,PRB71104402}

In SmFe$_4$P$_{12}$,
susceptibility and specific heat measurements show that the Sm ions are in a trivalent state
and that the system has an FM transition at the Curie temperature $T_{\text{C}}$ of 1.6 K.
\cite{JPCM15229,PB329458}
The temperature $T$ dependence of the electric resistivity exhibits the Kondo-lattice behavior
(Kondo temperature $T_{\text{K}} = 30$~K),
and the electronic specific heat coefficient attains the value of 370~mJ/mole K$^2$.
\cite{JPCM15229,PB329458}
In a BCC structure with the point group of $T_h$ symmetry,
the Sm$^{3+}$ ($J$ = 5/2) multiplet is split into a doublet $\Gamma _5$ and a quartet $\Gamma _{67}$
by the crystalline electric field (CEF).
The specific heat data also suggests
the CEF effect splitting of the ground state and the excited state separated by around 70 K.
\cite{JPSJ741030}  
However, the Kondo effect brings the uncertainty in the determination of the ground and the excited states,
$i.e.$, we do not know whether the ground state is $\Gamma _5$ or $\Gamma _{67}$.  
Indeed,
the magnetic entropy associated with the FM order is $0.17R\ln2$,
which is much smaller than the value expected for the case
that the ground state is $\Gamma _5$.

In this paper,
we report the microscopic magnetic properties of SmFe$_4$P$_{12}$
obtained by $^{31}$P-NMR (nuclear magnetic resonance) and $\mu$SR (muon spin relaxation) measurements. 
The results of the $^{31}$P-NMR Knight shift $K$ and the spin-lattice relaxation rate 1/$T_1$ measurements
are discussed on the basis of the Kondo screened magnetic moment model.
The $\mu$SR is ideal for the investigation of the magnetic properties in zero external field.

\section{Experimental}
The single-phase polycrystalline SmFe$_4$P$_{12}$ has been synthesized
by using the high temperature and high pressure method.
\cite{PB329458}
The sample was crushed into powder for the experiments.

The $^{31}$P-NMR has been performed
by using phase-coherent pulsed NMR spectrometers and superconducting magnets.
The $^{31}$P NMR spectra were measured by the convolution of Fourier transform signals of the spin echo
which were obtained at 20 kHz interval.
The 1/$T_1$ was measured with the saturation recovery method in applied magnetic fields of 0.921-33.4~kOe.
The nuclear magnetization recovery curve was fitted by a simple exponential function
as expected for the nuclear spin $I=1/2$ of $^{31}$P nucleus,
which allows us to determine a unique 1/$T_1$ value at each temperature and field. 

The $\mu$SR has been performed by implanting pulses of positive surface muons
at the RIKEN-RAL Muon Facility in the UK.
The direction of the initial muon spin is parallel to the beam line
where forward and backward counters were located on the upstream and the downstream sides.
The asymmetry parameter $A(t)$ was defined as
$A(t) = [F(t) - \alpha B(t)]/[F(t) + \alpha B(t)] - A_{\text{BG}}$,
where $F(t)$ and $B(t)$ are the total muon events counted
by the forward and the backward counters at time $t$, respectively.
The $A_{\text{BG}}$ is the background,
and $\alpha$ is the calibration factor reflecting the relative counting efficiencies of both counters.
In zero field (ZF-) experiments,
stray fields at the sample position were compensated within 0.03~Oe by using correction coils,
which is small enough for our ZF-$\mu$SR measurement.

\section{Results and Discussion}
\begin{figure}[t]
\begin{center}
\includegraphics[width=8.6cm]{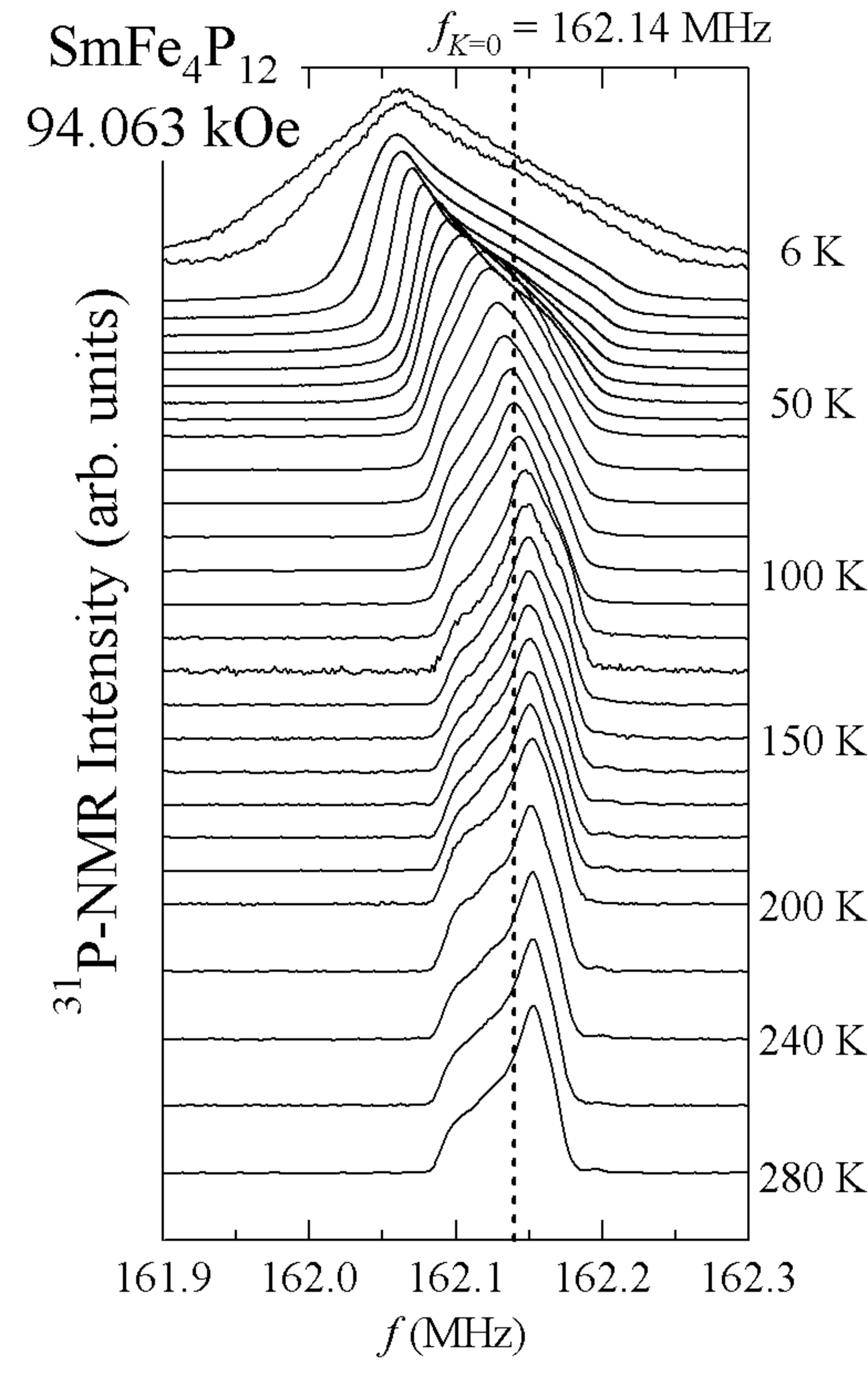}
\end{center}
\caption{
$^{31}$P-NMR spectra at several temperatures in the applied magnetic field of 94.063 kOe.
The broken line shows the resonance frequency of 162.14 MHz at $K = 0$
estimated from the nuclear gyromagnetic ratio of the $^{31}$P nucleus.
}
\label{f1}
\end{figure}

Figure \ref{f1} shows the $^{31}$P-NMR spectra at several temperatures in the applied field of 94.063~kOe.
The line shape of the spectra at high temperatures shows a powder pattern
with a nearly uniaxial Knight shift distribution for nuclear spin $I= 1/2$.
However,
the line shape gradually changes and deviates from the uniaxial powder pattern
with decreasing temperature.
The change of the line shape becomes remarkable below 70~K.
The spectra were analyzed
by taking into account of the powder average of $K$ anisotropy and the excess Gaussian broadening.
The $K$ is expressed as
\begin{equation}
K = K_\text{iso} + K_1 (3 \cos^2 \theta - 1) - K_2 (\sin^2 \theta \cos 2 \phi),\\
\end{equation}
where the $K_{\text{iso}}$ represents an isotropic term,
$K_{1}$ represents a uniaxial term, and $K_{2}$ represents the deviation from the uniaxial symmetry. 

\begin{figure}[t]
\begin{center}
\includegraphics[width=8.6cm]{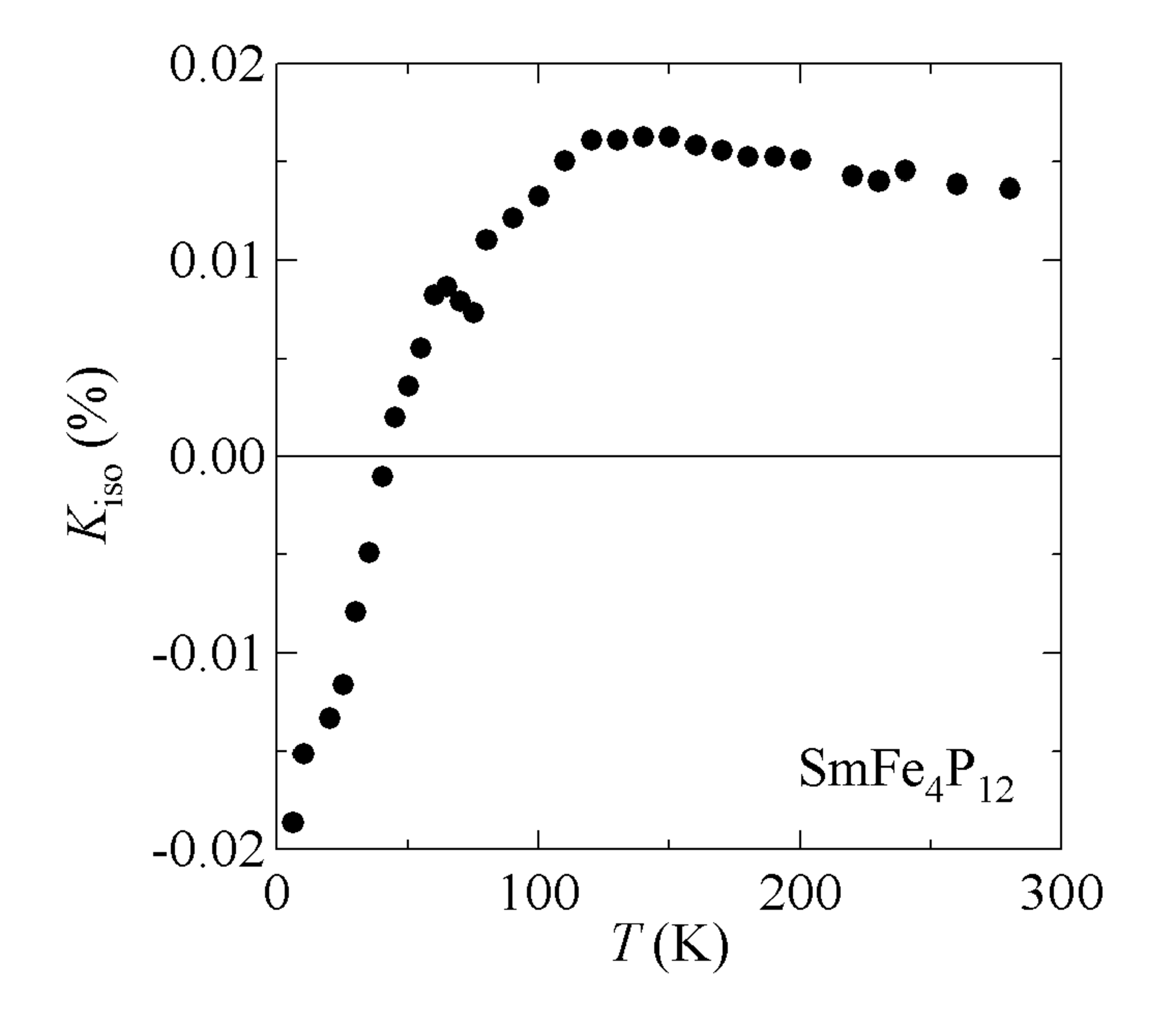}
\end{center}
\caption{
Temperature dependences of $K_{\text{iso}}$
}
\label{f2}
\end{figure}

\begin{figure}[t]
\begin{center}
\includegraphics[width=8.6cm]{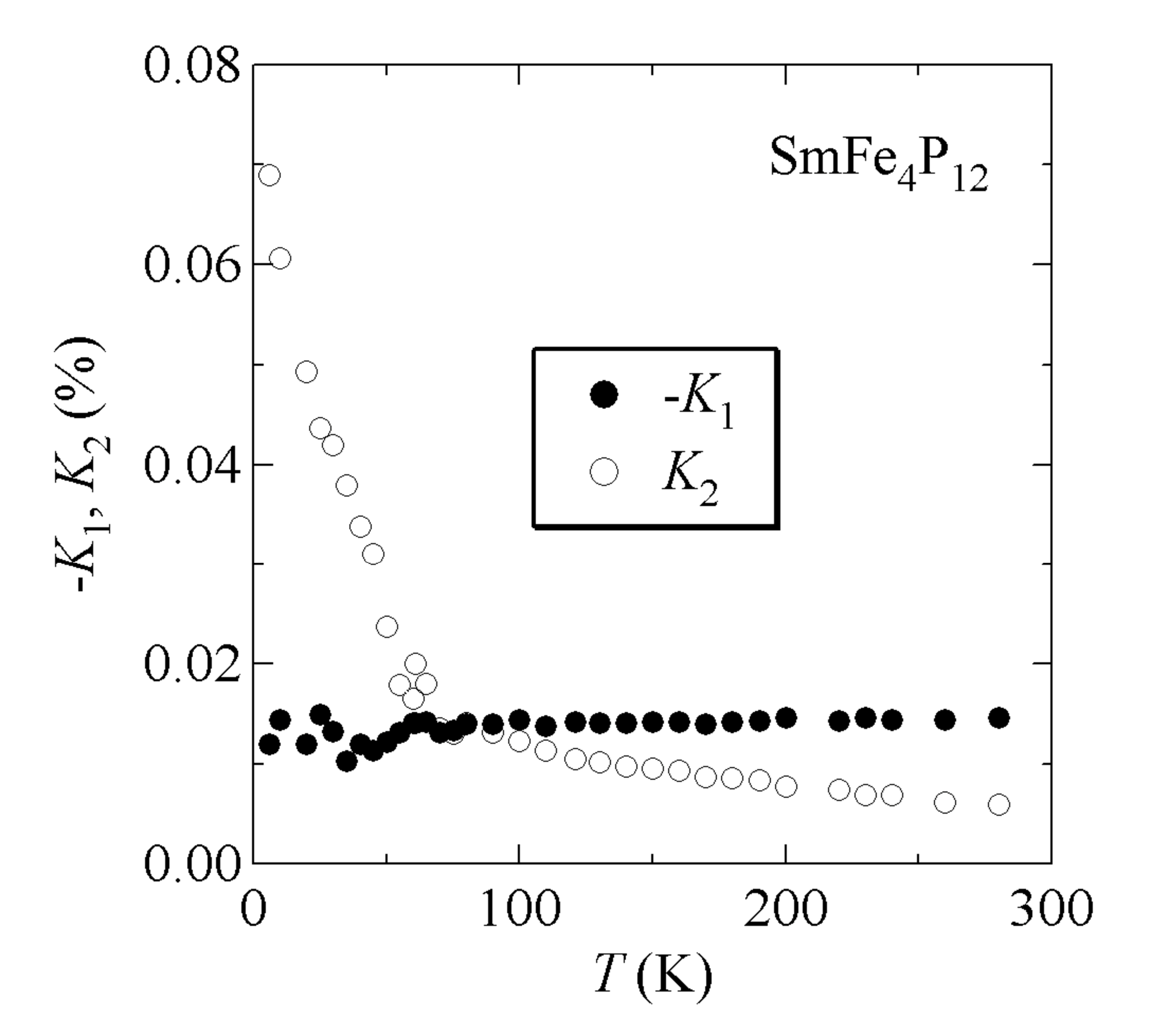}
\end{center}
\caption{
Temperature dependence of $-K_1$ (closed circles) and $K_2$ (open circles).
}
\label{f3}
\end{figure}

\begin{figure}[t]
\begin{center}
\includegraphics[width=8.6cm]{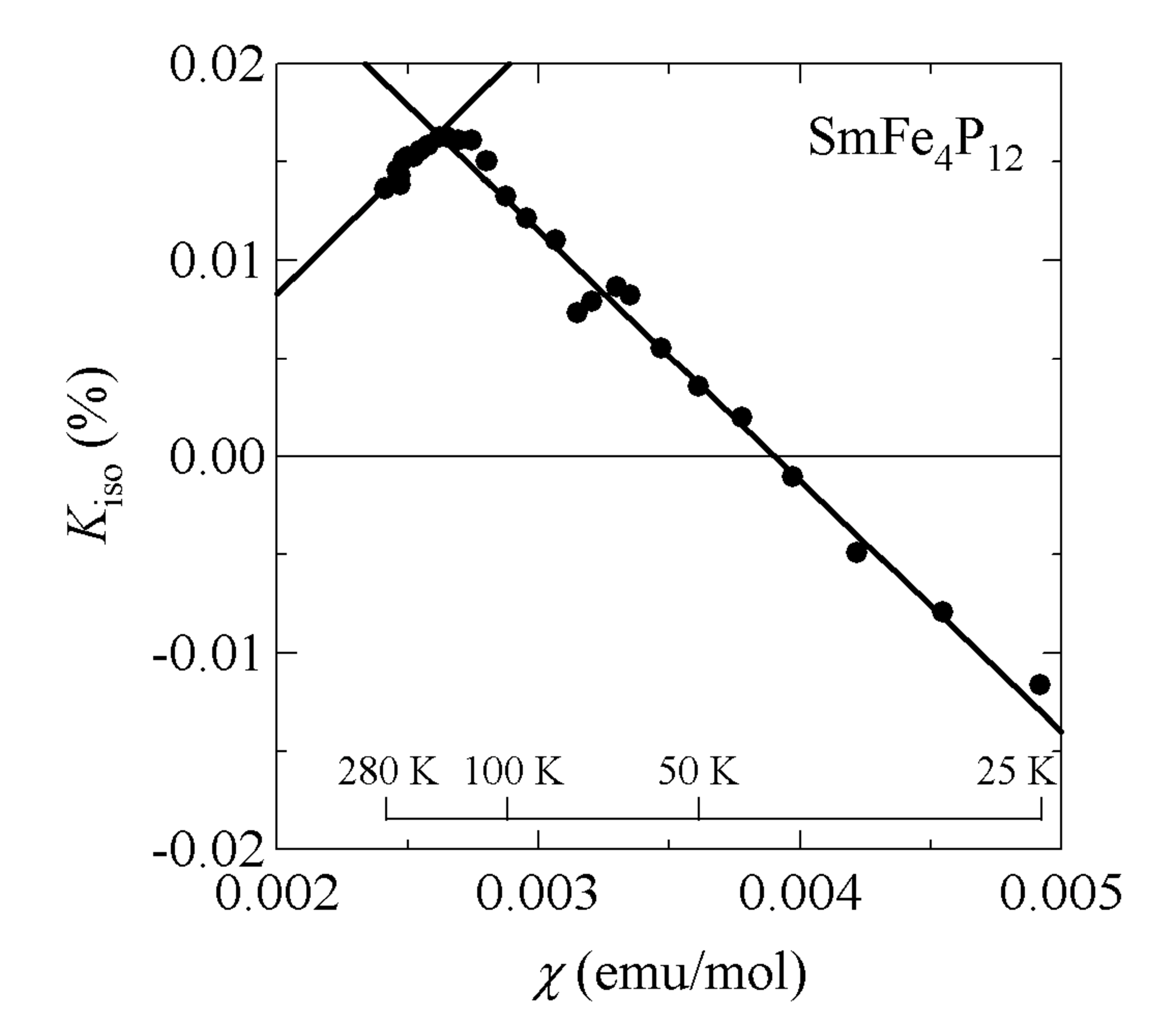}
\end{center}
\caption{
Bulk susceptibility dependence of $K_{\text{iso}}$ ($K$-$\chi$ plot).
The solid lines show the linear function of $\chi$ expressed by the formula (\ref{K-chi}).
Some temperatures are also shown corresponding to $\chi$.
}
\label{f4}
\end{figure}

The $T$ dependence of $K_{\text{iso}}$, $-K_1$ and $K_2$ are shown in Figs. \ref{f2} and \ref{f3}, respectively.
The $T$ dependence of $K_{\text{iso}}$ is remarkably different from that of the bulk susceptibility $\chi$.
The $K_{\text{iso}}$ rapidly increases with increasing temperature,
and has a broad maximum around 120 K and then decreases slightly at higher temperatures.
To clarify the $T$ dependence,
$K_{\text{iso}}$ is plotted against $\chi$, as seen in Fig. \ref{f4}.
The $K$-$\chi$ plot shows the $T$ dependence of the transferred hyperfine coupling as 
\begin{equation}
K_{\text{iso}} = \frac{A_{\text{hf}}}{N_{\text{A}} \mu_{\text{B}}} \chi \label{K-chi},
\end{equation}
where $N_{\text{A}}$ and $\mu_{\text{B}}$ are the Avogadro number and the Bohr magneton, respectively.
It is noticed that the $K$-$\chi$ plot is linear in the $T$ ranges of 25-120~K and above about 150~K. 
The hyperfine coupling constants $A_{\mathrm{hf}}$ were estimated
about $- 0.71$ kOe/$\mu _{\mathrm{B}}$ below 120~K
and about $+ 0.74$ kOe/$\mu _{\mathrm{B}}$ above 150~K
from the $K$-$\chi$ plot by adopting the formula (\ref{K-chi}).
Considering $T_{\text{K}} = 30$ K in this system,
the change in the hyperfine coupling around 120~K is reasonably attributed to the CEF splitting.
The change of a hyperfine field associated with a CEF effect was reported in other system.
\cite{JPSJ642628}

The $K_1$ is nearly $T$ independent in the whole $T$ region with the value of about $-0.015$~\%. 
The $K_2$ decreases with increasing temperature below about 70~K,
and decreases very gradually at higher temperatures ($0.0059$~\% at 280~K).
One origin of the anisotropic hyperfine coupling is the the direct dipolar coupling
of Sm$^{3+}$ magnetic moments and $^{31}$P nuclear spins. 
The lattice sum calculation predicts nearly uniaxial Knight shift to be
$K_1 \simeq 0.014$~\% and $K_2 \simeq 0.0017$~\% at 280 K.
These results are shown in Table \ref{t1} including the experimental results.
Though the uniaxial anisotropy is observed in our system at high temperatures,
$K_1$ has the opposite sign.
This result indicates that
the hyperfine mechanism can not be explained by only the direct dipolar coupling
even at sufficiently high temperatures (280 K).
The nearly same magnitude of $K_2$ at high temperatures
between the experimental and the calculated values is accidental.
This suggests that another hyperfine field mechanism contributes to the anisotropic Knight shift. 
The hybridization of the Sm-4f orbital with the P-3p orbital makes spin polarizations at P-3p spins,
which induces the anisotropic field via the dipolar coupling of the P-3p spins and $^{31}$P nuclei. 
The sum of these contributions would explain the observed values.
The change of the spectra below 70 K is
represented by the increase of $K_2$ which reflects the CEF splitting of about 70 K
\cite{JPSJ741030}.
The change of a Sm-4f orbital shape occurs associated with the thermal excitation of the CEF levels 
and modifies the hybridization of the Sm-4f and the ligand P orbitals.
For the evaluatation of the ligand spin polarization,
a complete solution of the hybridization process is required.
The analysis of the spectra is an open problem for future works.   

\begin{table}[t]
\caption{
Experimental and the dipolar calculated values of $K_1$ and $K_2$ at the temperature of 280 K.
}
\label{t1}
\begin{center}
\begin{tabular}{c|rr}
\hline\hline
&$K_1$ (\%)&$K_2$ (\%)\\
\hline
Experimental&$-0.015$&$0.0059$\\
Calculated&$0.014$&$0.0017$\\
\hline\hline
\end{tabular}
\end{center}
\end{table}

\begin{figure}[t]
\begin{center}
\includegraphics[width=9cm]{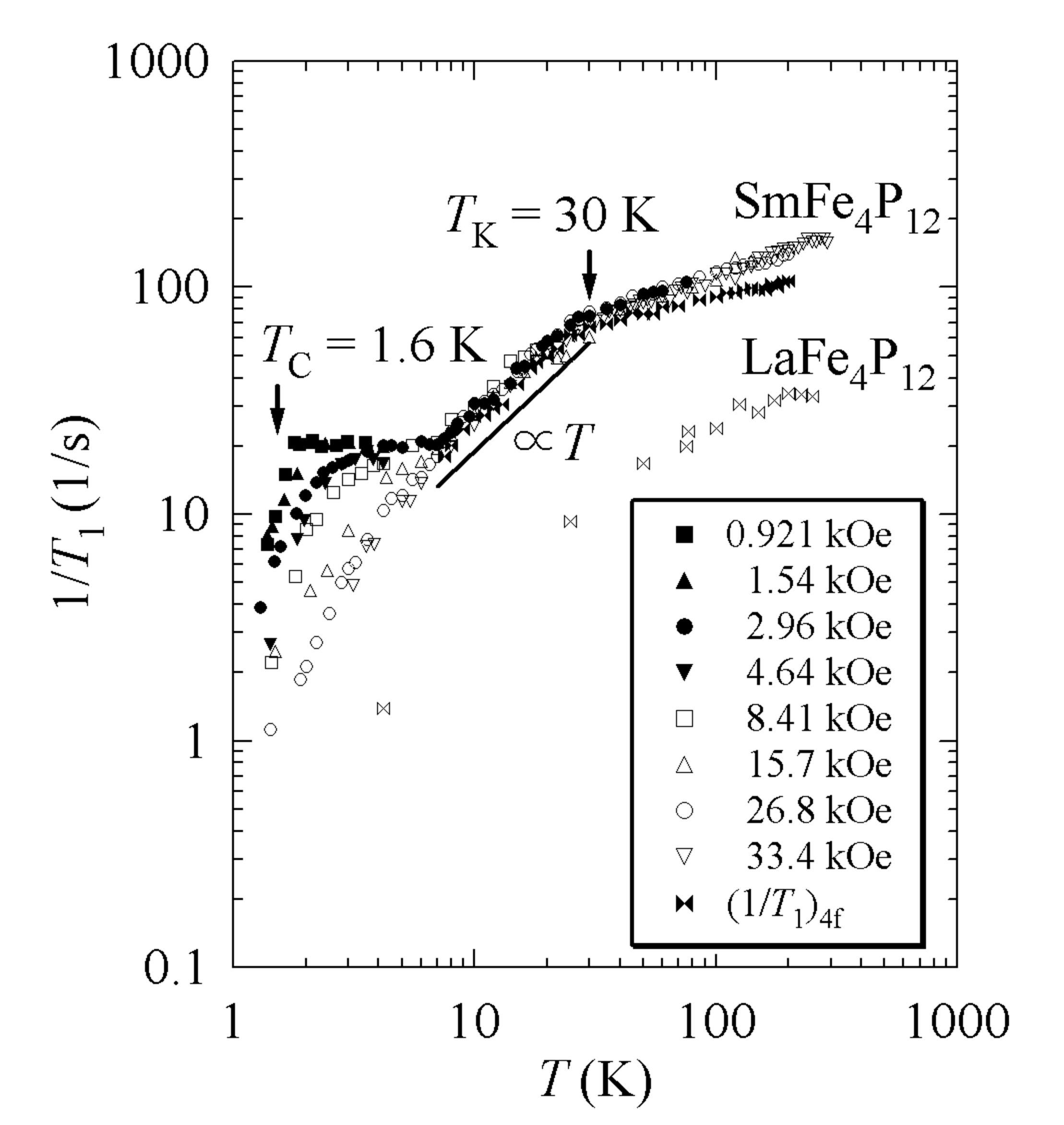}
\end{center}
\caption{
Temperature dependence of 1/$T_1$ in several applied magnetic fields.
The $1/T_1$ of nonmagnetic LaFe$_4$P$_{12}$ is also shown as a reference.
\cite{JPSJ743370}
``$(1/T_1)_{\text{4f}}$'' is obtained by
subtracting $1/T_1$ of LaFe$_4$P$_{12}$ from that of SmFe$_4$P$_{12}$.
}
\label{f5}
\end{figure}

Figure \ref{f5} shows the $T$ dependence of $1/T_1$ in several applied fields.
The $T_1$ tends to become long with decreasing temperature.
Above 30~K, the weak $T$ dependence was observed.
This is attributed to the relaxation by the 4f localized moments of the Sm ions.
We evaluated the contribution of the conduction electrons by $1/T_1$ of the nonmagnetic LaFe$_4$P$_{12}$,
which is much smaller than the value observed in SmFe$_4$P$_{12}$.
\cite{JPSJ743370}  
After subtracting this contribution,
the corrected 1/$T_1$ has much weaker $T$ dependence,
which is shown as ``$(1/T_1)_{\text{4f}}$'' in Fig. \ref{f5}.
We observed the Fermi liquid (the Korringa behavior, $T_1T \sim \text{constant}$)
in the temperature range of 7.5-30 K,
which indicates the major part of the Sm 4f-electrons to be itinerant
by the hybridization of the conduction electrons below $T_{\text{K}}$.

The $1/T_1$ at 0.921 kOe is nearly $T$ independent above $T_{\mathrm{C}}$,
and rapidly decreases at lower temperatures. 
The rapid decrease suggests the existence of a phase transition in ZF.
The critical slowing-down phenomenon of $1/T_1$ might be suppressed even by the small external field.
We estimated $1/T_1$ on the presumption
that the Kondo screened Sm$^{3+}$ moments $\sim$0.2$\mu_{\text{B}}$
coupled by the ferromagnetic exchange interactions.
\cite{JPCM15229}
The 1/$T_1$ is expressed as
\begin{equation}
\frac{1}{T_1} = \frac{\sqrt{2 \pi} J (J+1) Z(A/Z)^2}{3 \hbar^2 \omega_e} \simeq \frac{\omega_n^2}{\omega_e},
\end{equation}
where
\begin{gather}
\omega_n = \frac{AJ}{\hbar} = \gamma H_{\text{int}}, \quad H_{\text{int}}= H_{\text{hf}} \mu,\\
\omega_e = \frac{2}{3} \frac{J_{\text{exc}}^2}{\hbar^2} ZJ(J+1), \quad J_{\text{exc}} = \frac{3 k_{\text{B}} T_{\text{C}}}{2ZJ(J+1)},
\end{gather}
$J$ is the quantity of the angular momentum of the Sm ions,
$Z$ is the number of the nearest neighbor Sm ions,
$\hbar$ is the Planck constant divided by 2$\pi$,
$\gamma$ is the nuclear gyromagnetic ratio of the $^{31}$P nucleus,
$H_{\text{int}}$ is the internal field made by the 4f localized moments of the Sm ions at the $^{31}$P spins,
$\mu$ is the saturated moment at 1.4~K,
$J_{\text{exc}}$ is the ferromagnetic exchange interaction energy,
and $k_{\text{B}}$ is the Bolzmann constant.
The estimated value of $1/T_1$ is about 11.6 1/s,
which is close to the experimental values.
This result suggests that the origin of the ferromagnetic ground state would be due to
the uncovered 4f local moments of the Sm$^{3+}$ by the conduction electrons.
The value of $1/T_1$ at low $T$ decreases rapidly with increasing field.
The coupling of the uniform ferromagnetic fluctuations and external field strongly suppresses 1/$T_1$.

It is important to carry out ZF-measurements to
identify the anomaly around $T_{\text{C}}$ as a phase transition.
However,
it is impossible to perform $^{31}$P-NMR in ZF,
since the $^{31}$P nucleus ($I$=1/2) has no quadrupole moment.
Hence, we carried out ZF-$\mu$SR to confirm the phase transition.

\begin{figure}[t]
\begin{center}
\includegraphics[width=8.6cm]{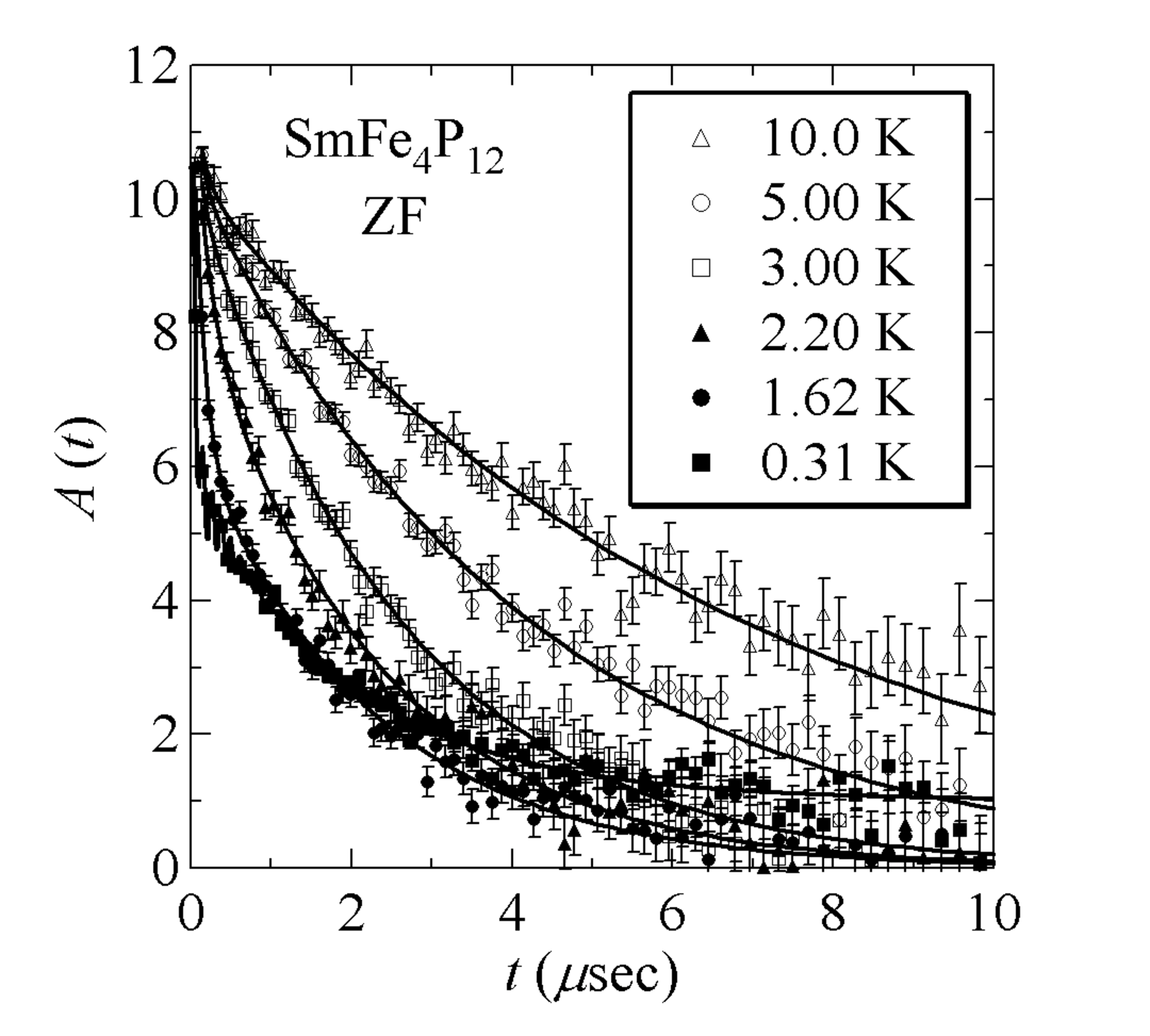}
\end{center}
\caption{
ZF-$\mu$SR time spectra at several temperatures.
The solid lines show the best-fit results by the formula (\ref{Dbexp_precession}).}
\label{f6}
\end{figure}

\begin{figure}[t]
\begin{center}
\includegraphics[width=8.6cm]{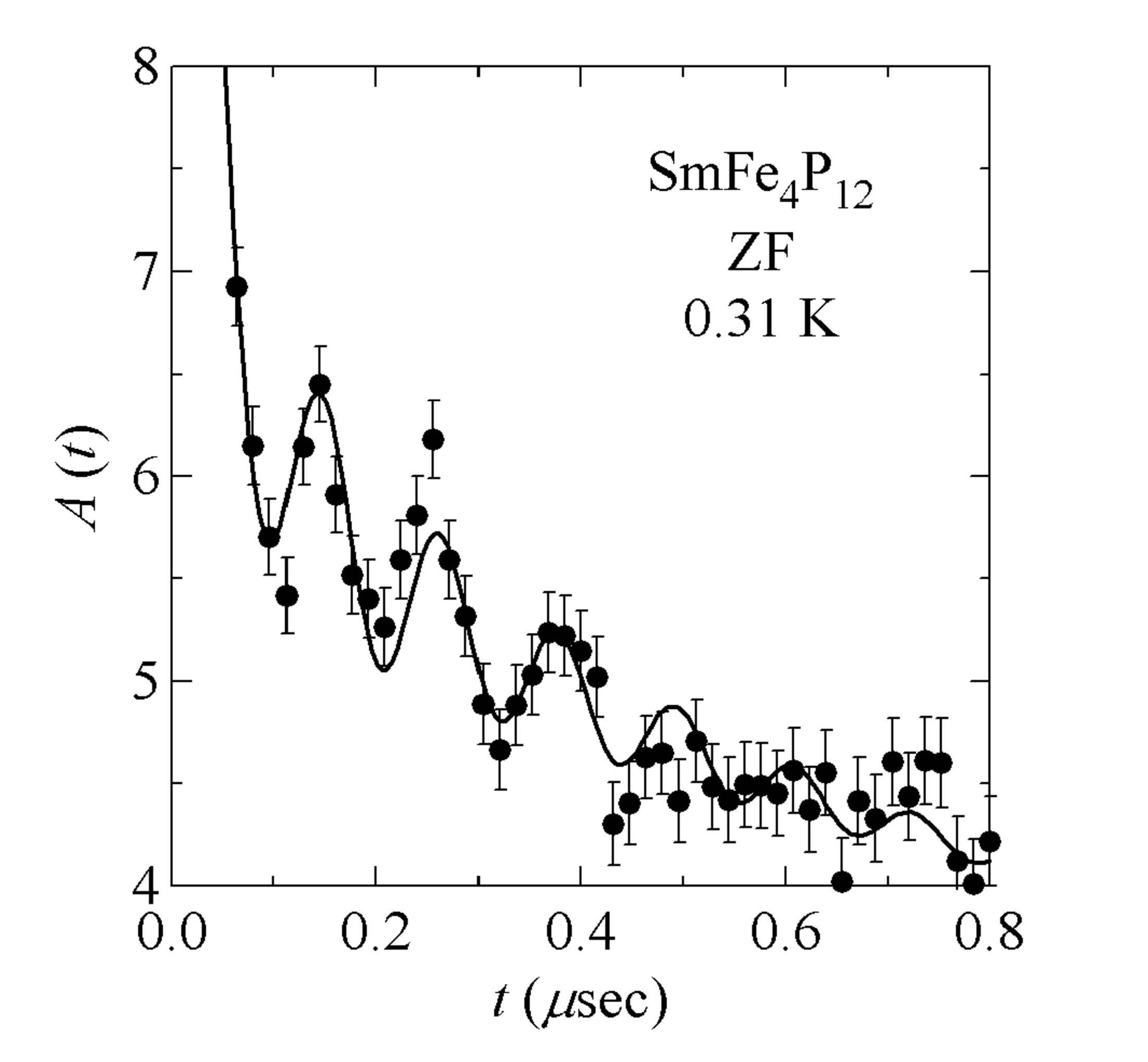}
\end{center}
\caption{
ZF-$\mu$SR time spectrum at the temperature of 0.31~K,
which is obtained at a high time resolution.
The solid line shows the best-fit results by the formula (\ref{Dbexp_precession}).}
\label{f7}
\end{figure}

Figure~\ref{f6} shows the ZF-$\mu$SR time spectra at several temperatures.
Below $T_{\text{C}}$,
$A(t)$ rapidly decreases,
and the muon-spin precession was observed as seen clearly in Fig~\ref{f7}.
This indicates that the system has a magnetically ordered ground state.
The internal field of about 650~Oe at the muon site was estimated
from the precession frequency by the formula~(\ref{Dbexp_precession}).
Above $T_{\text{C}}$,
$A(t)$ gradually decreases with $t$ 
where muon-spins depolarize slowly by transferred magnetic fields
from dynamically fluctuating Sm moments and nuclear moments.

\begin{figure}[t]
\begin{center}
\includegraphics[width=8.6cm]{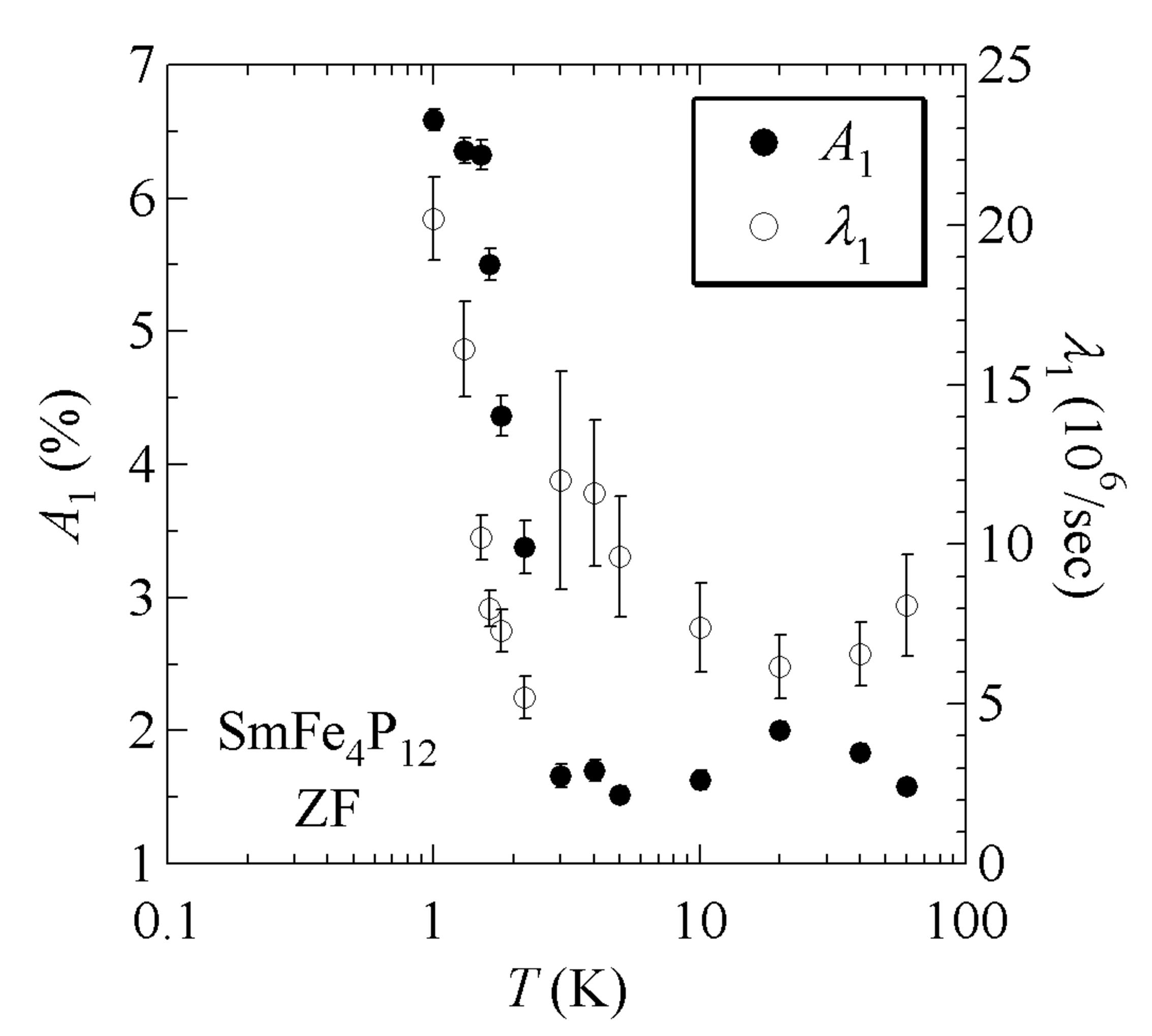}
\end{center}
\caption{
Temperature dependence of $A_1$ (closed circles, left axis) and $\lambda_1$ (open circles, right axis)
above 1~K obtained from the best-fit results shown as the solid lines in Fig. \ref{f6}.
}
\label{f8}
\end{figure}

\begin{figure}[t]
\begin{center}
\includegraphics[width=8.6cm]{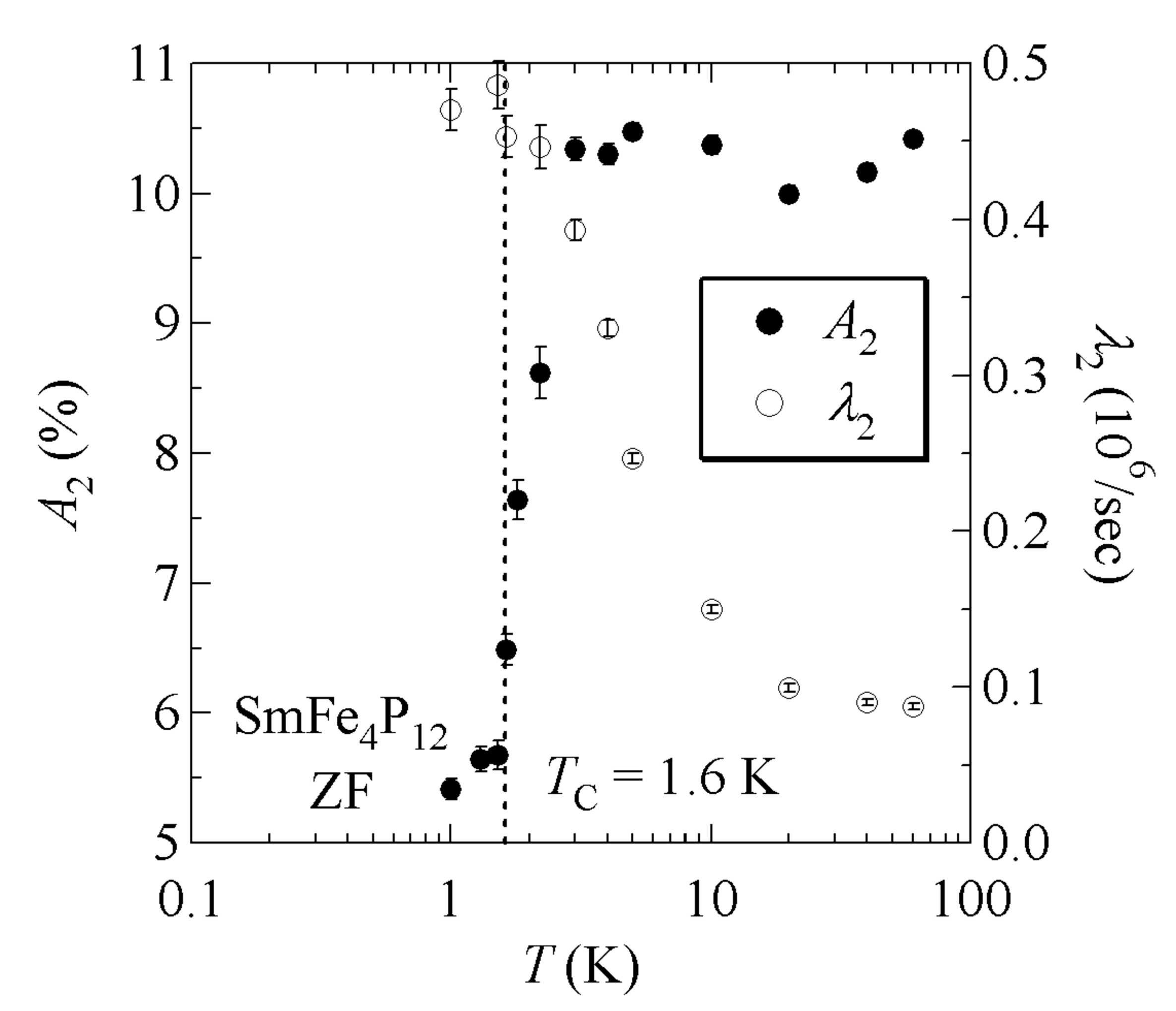}
\end{center}
\caption{
Temperature dependence of $A_2$ (closed circles, left axis) and $\lambda_2$ (open circles, right axis)
above 1~K obtained from the best-fit results shown as the solid lines in Fig. \ref{f6}.
}
\label{f9}
\end{figure}

The spectra were analyzed by a multi-components function expressed by the formula
\begin{equation}
A(t) = A_1e^{-\lambda_1 t} + A_2e^{-\lambda_2 t} + A_3e^{-\lambda_3 t} \cos (\omega t + \theta).
\label{Dbexp_precession}
\end{equation}
The first and the second terms simply represent
the components of the rapid and the slow depolarizations, respectively.
The third term describes the precession component.
The parameters of $A_1$, $A_2$, $A_3$ are the initial asymmetries at $t=0$,
the $\lambda_1$, $\lambda_2$, $\lambda_3$ are the muon-spin depolarization rates,
and $\omega$, $\theta$ are the frequency and the phase of the precession, respectively.
The spectra above 1~K which have no precession were analyzed without the third term.
Figures~\ref{f8} and \ref{f9} show
the $T$ dependence of $A_1$, $\lambda_1$ and $A_2$, $\lambda_2$ above 1~K
obtained from the analysis, respectively.
It is noted that the sum of $A_1$ and $A_2$ is the constant value of $A(0)$.
The spectra above 3~K are almost dominated by only the second term as seen in $A_1$ and $A_2$
where the spectra almost follow a single exponential function.
Hence, 
the absolute values of $A_1$ (corresponding to $A_2$) and $\lambda_1$ above 3~K are not so intrinsic,
and the value of $A_1$ should be 0 idealy.
Below 3~K,
the first term (rapid depolalization) represents
the relaxation by both static and dynamical components
which develop below the temperature.
Then, we forcus our attention on $\lambda_2$ of the second term
to discuss the dynamical component.

As seen in Fig.~\ref{f9},
clear anomalies were observed in both $A_2$ and $\lambda_2$ around $T_{\text{C}}$.
The rapid decrease of $A_2$ corresponds to the rapid increase of $A_1$.
The appearance of the fast relaxation behavior suggests
the development of a static internal field attributed to the existence of a magnetic order.
Considering that the slow relaxation of the spectra is due to a dynamical component, 
the peak in $\lambda_2$ arises from the critical slowing down behavior of the Sm moments around $T_{\text{C}}$.
The observation of the critical slowing down behavior seems to contradict that of $1/T_1$ by $^{31}$P-NMR.
One reason would be due to the time-window difference of each probe.
The gyromagnetic ratio of the muon is two orders of the magnitude larger than that of $^{31}$P nucleus.
Hence, $\mu$SR is much sensitive to higher frequency components of magnetic fluctuations,
which makes the broad temperature dependence of $\lambda_2$.	
It is noteworthy that the behavior of $\lambda_2$ is not due to the inhomogeneity of the sample
because the sample used in this measurement had been well characterized by other macroscopic measurements.

Though the bahavior of $A_2$ and $\lambda_2$ is obtained qualitatively,
this is consistent with the results of $^{31}$P-NMR and other measurements.
\cite{PB329458,JPCM15229,JPSJ741030}
Indeed,
these behavior was also observed around $T_{\text{MI}}$ in SmRu$_4$P$_{12}$
where the magnetic order below $T_{\text{MI}}$ has been confirmed by previous $\mu$SR measurements.
\cite{PRB73052408}
The most important point of this ZF-$\mu$SR result of SmFe$_4$P$_{12}$ is
the observation of the muon-spin precession below $T_{\text{C}}$.
These results conclude that the magnetic order occurs below $T_{\text{C}}$,
and support the $^{31}$P-NMR results which could not be obtained in ZF.

The FM fluctuations around $T_{\text{C}}$ in finite fields
and the existence of a magnetic phase transition at $T_{\text{C}}$ in ZF
were observed by means of $^{31}$P-NMR and $\mu$SR, respectively.
Therefore,
an FM phase transition at $T_{\text{C}}$ in SmFe$_4$P$_{12}$ has been confirmed.

\section{Conclusions}
The $^{31}$P-NMR in external fields and the ZF-$\mu$SR studies on SmFe$_4$P$_{12}$
have been carried out.
These results show the existence of the CEF effect,
the HF behavior below $T_{\text{K}}$ of 30~K
and the FM order below $T_{\text{C}}$ of 1.6~K.

\section*{Acknowledgments}
The authors would like to thank K. Matsuhira and T. Hotta for their useful discussions.
This work was supported by a Grant-in-Aid for Scientific Research
from the Ministry of Education, Sport, Science and Culture of Japan (no. 17740212).
The $\mu$SR work was supported by the Toray Science Foundation
and the Joint Project of Japan Society for the Promotion of Science.
The work at Muroran Institute of Technology was supported by
a Grant-in-Aid for Scientific Research in Priority Area ``Skutterudite'' (no. 15072201)
of the Ministry of Education, Culture, Sports, Science and Technology, Japan.

\end{document}